\documentstyle[amssymb,12pt]{article}

\begin{document}
\title{RECURSION OPERATORS OF SOME EQUATIONS OF 
HYDRODYNAMIC TYPE}
\author{M. G{\" u}rses and K. Zheltukhin\\
{\small Department of Mathematics, Faculty of Sciences}\\
{\small Bilkent University, 06533 Ankara-Turkey}\\
{\it email: gurses@fen.bilkent.edu.tr}}
\begin{titlepage}
\maketitle
\begin{abstract}
We give a general method for constructing  recursion operators for
some equations of hydrodynamic type, admitting a nonstandard Lax
representation.
We give several examples for $N=2$ and $N=3$ containing the equations of
shallow water waves and its generalizations with their first two higher
symmetries and their recursion operators. We also discuss a reduction
of $N+1$ systems to $N$ systems of some new equations of hydrodynamic
type.

\end{abstract}
\end{titlepage}

\section*{I. Introduction}

Most of the integrable nonlinear partial differential equations admit Lax
representations

\begin{equation}
L_t=[A,L],
\end{equation}

\noindent
where $L$ is a pseudo-differential operator of order $m$ and $A$ is
 a pseudo differential operator. Recently \cite{GKS} we  established 
a new method
for such integrable equations to construct their recursion operators.
This method uses the hierarchy of equations  

\begin{equation} 
L_{t_n}=[A_n,L]
\end{equation}

\noindent
and the Gel'fand-Dikkii \cite{GelDik} construction of the $A_n$-operators.
Defining an operator $R_n$ in the form

\begin{equation}
A_n=LA_{n-m} +R_n, \label{anm}
\end{equation}

\noindent
one  then obtains relations among the hierarchies 

\begin{equation}
L_{t_n}=LL_{t_{n-m}} + [R_n;L]
\end{equation}

\noindent
This equation allows to find  $L_{t_n}$ in terms of $L_{t_{n-m}}$.
It is important to note that one does not need  to know the exact 
form of $A_n$. For further details of the method see \cite{GKS}.

Here we extend this method to equations of hydrodynamic type \cite{gur2}. 
These equations
and their Hamiltonian formulation (sometimes called the dispersion-less KdV 
system) were studied by  Dubrovin and Novikov
\cite{nd}. See \cite{frpt} for more details on this subject. It is known
that these equations admit a nonstandard Lax representation

\begin{equation}
\frac{\partial L}{\partial t}=\{A,L\}_{k},
\end{equation}

\noindent
where $A,L$ are differentiable functions of $t,x,p$ on a Poisson manifold $M$ 
with local coordinates $(x,p)$ and $\{,\}_{k}$ is the Poisson bracket. 
On $M$ we take this Poisson bracket $\{ , \}_{k}=
p^{k}\,\{ , \}$, where $\{ , \}$ is the canonical Poisson bracket and k is 
an integer. For more information on Poisson manifolds see \cite{Olv},\cite{PG}.
Equations of hydrodynamic type with the above Lax representations
were studied in \cite{StrF}-\cite{BGZ}.
Having such Lax representation, we  can  consider a whole hierarchy of 
equations

\begin{equation}
  \label {LaxA}
\frac{\partial L}{\partial t_n}= \{A_n,L\}_{k}. 
\end{equation}

\noindent
We can also represent function $A_{n+m}$ in the form given in (\ref{anm})
and apply our  method \cite{GKS} for constructing of a recursion operator 
for the equation (\ref{LaxA}).  There are some other works \cite{sheft}-
\cite{fg} which give also recursion operators of some equations of
hydrodynamic type. The form of these operators are different then
the recursion operators presented in this work. Our method \cite{GKS}
produces recursion operators for hydrodynamic type of equations in
the form ${\cal R}=A+B\,D^{-1}$ where $A$ and $B$ are functions of 
dynamical variables and their derivatives. All higher symmetries
obtained by the repeated application of this recursion operator to 
translational symmetries belong also to the hydrodynamic type
of equations. The recursion operators obtained in references
 \cite{sheft}-\cite{fg} are of the form ${\cal R}=C\,D+A+B\,D^{-1}\,E$,
where $A,B,C$, and $E$ are functions of dynamical variables and their 
derivatives.

In the next section we discuss the Lax representation with Poisson
brackets for polynomial Lax functions.
In Sec.III we give the method of construction of the recursion operators
following \cite{GKS}. In Sec.IV  we give several examples for
$k=0$ and $k=1$. In Sec.V we consider the Poisson bracket for general
$k$ and let

\begin{equation}
L=p+S+Pp^{-1} ,
\end{equation}

\noindent
and find the Lax equations and the corresponding recursion operator
for $N=2$. In Sec.VI we consider the Lax function

\begin{equation}
L=p^{\gamma-1}+u+{v^{\gamma-1} \over (\gamma-1)^2}\, p^{-\gamma+1},
\end{equation}

\noindent
and take $k=0$. We obtain the equations corresponding to the
polytropic gas dynamics and its recursion operators \cite{Olv},
\cite{BurD}. It is interesting to note
that   the systems of equations and their recursion operators obtained 
in these sections
V and VI are transformable into each other. In Sec.VII  we give a method
 reduction
from $N+1$ system to an $N$ system by letting one of the symmetrical
variables (defined in the text) to zero. Reduced systems are shown to
be also integrable, i.e., they admit recursion operators.

\section*{II. Lax formulation with Poisson bracket}

We start with the definition of the standard Poisson bracket. Let
$f(x,p)$ and $g(x,p)$ be differentiable functions of their 
arguments. Then the standard Poisson bracket is defined by (see
\cite{Olv} and \cite{Str} for more details)

\begin{equation}
\{f,g\}={\partial f \over \partial p}\, {\partial g \over \partial x}-
{\partial f \over \partial x}\,{\partial g \over \partial p}.
\end{equation}

\noindent
We give a slight modification of this bracket as \cite{Str}

\begin{equation}
\{f,h\}_{k}=p^{k}\, \{f,g\},
\end{equation}

\noindent
where $k$ is an integer. Then we have

\vspace{0.3cm}

\noindent
{\bf Lemma 1}. {\it For any $k\in {\Bbb Z} $ the bracket $\{f,g\}_k$
defines  a Poisson brackets.}

\vspace{0.3cm}

\noindent
{\bf Proof.} 
We should check only the Jacobi identity.  Other properties of Poisson
brackets are evidently true.
The standard bracket  $\{f,g\}$  satisfies the Jacobi identity. For
all $k$ we have to show that 

$$
\{\{f,g\}_k,h\}_k + \{\{h,f\}_k,g\}_k + \{\{g,h\}_k,f\}_k=0. 
$$

\noindent
First, note that 

$$
\{\{f,g\}_k,h\}_k=p^k\{p^k\{f,g\},h\}=
p^{2k}\{\{f,g\},h\} + kp^{k-1}\{f,g\}h_x. 
$$

\noindent
Thus, we have

$$
\{\{f,g\}_k,h\}_k + \{\{h,f\}_k,g\}_k + \{\{g,h\}_k,f\}_k  = 
                    p^{2k}(\{\{f,g\},h\} + \{\{h,f\},g\} +   
$$
$$
 \qquad \{\{g,h\},f\}) +  kp^{k-1}( \{f,g\}h_x + \{h,f\}g_x + \{g,h\}f_x ).  
$$

\noindent
Equality

$$
\{\{f,g\},h\} + \{\{h,f\},g\} + \{\{g,h\},f\}=0 
$$

\noindent
holds and it is easy to check that

$$
\{f,g\}h_x + \{h,f\}g_x + \{g,h\}f_x =0.
$$

\noindent
Hence, $\{,\}_{k}$, for all $k$  defines a Poisson bracket.
$\Box$

\medskip

For each $k\in {\bf Z}$ we can consider  hierarchies of equations of 
hydrodynamic type, defined in terms of the Lax function 

\begin{equation}
 \label{Laxfnc}
L= p^{N-1} + \sum_{i=-1}^{N-2} p^i S_i(x,t)
\end{equation}

\noindent
by the Lax equation

\begin{equation} 
 \label{Laxeqn}
\frac{\partial L}{\partial t_n}=
         \left\{ ( L^{\frac{n}{N-1}} )_{\ge -k+1};L \right\}_k, 
\end{equation}

\noindent
where $n=j+l(N-1)$ and  $j=1,2, \dots,(N-1),\; l\in {\bf N}$ . So
we have a hierarchy for each $k$ and $j=1,\dots, (N-1)$. 
Also, we require $n \ge -k+1$ to ensure that 
$(L^{\frac{n}{N-1}})_{\ge -k+1}$ is not zero.
With the choice of Poisson  
brackets $\{,\}_{k}$, we must take a certain part of the series expansion of
 $L^{\frac{n}{N-1}}$ to get the consistent equation (\ref{Laxeqn}).
This part is  $( L^{\frac{n}{N-1}} )_{\ge -k+1}$.

The Lax function (\ref{Laxfnc}) can also be written in terms of symmetric
variables $u_1, \dots, u_N$

\begin{equation}
L= \frac{1}{p}\prod_{j=1}^{N}(p-u_j)
\end{equation}

\noindent
that is $u_1, \dots, u_N$ are roots of the polynomial 

$$
p^{N-1} + S_{N-2}p^{N-2 } + \dots +S_{-1}. 
$$

\noindent
In new variables the equation (\ref{Laxeqn}) is invariant under
 transposition of variables.

\section*{III. Recursion Operators}

For each hierarchy of the equations (\ref{Laxeqn}), depending on the pair
 $(N,k)$, we can find a recursion operator.

\vspace{0.3cm}

\noindent
{\bf Lemma 2}. {\it For any $n$

\begin{equation}       
\label{SymmRel}
L_n=LL_{n-(N-1)} + \{R_n;L\}_k,
\end{equation}

\noindent
where function $R_n$ has a form

\begin{equation}
\label{FormRem}
R_n= \sum_{i=0}^{N-2} p^{i-k}A_i(S_{-1}\dots S_{N-2}, S_{-1,n-1}\dots
 S_{N-2,n-1}).
\end{equation}
}

\vspace{0.3cm}

\noindent
{\bf Proof.}

$$
( L^{\frac{n}{N-1}} )_{\ge -k+1} = [L(L^{\frac{n}{N-1}-1})_{\ge-k+1} +
                        L(L^{\frac{n}{N-1}-1})_{<-k+1}]_{\ge -k+1} 
$$

\noindent
So,

\begin{eqnarray}
( L^{\frac{n}{N-1}} )_{\ge -k+1} = L( L^{\frac{n}{N-1}-1} )_{\ge-k+1} \: +
(L( L^{\frac{n}{N-1}-1} )_{<-k+1})_{\ge -k +1} \nonumber \\
 - (L( L^{\frac{n}{N-1}-1} )_{\ge -k+1})_{<-k+1} . 
\end{eqnarray}

\noindent
If we put

$$
R_n=(L( L^{\frac{n}{N-1}-1} )_{<-k+1})_{\ge -k +1} \:
     - (L( L^{\frac{n}{N-1}-1} )_{\ge -k+1})_{<-k+1} \quad , 
$$

\noindent
then 

$$
( L^{\frac{n}{N-1}} )_{\ge -k+1} = L( L^{\frac{n}{N-1}-1} )_{\ge
  -k+1}\: + R_n .
$$

\noindent
Hence,

\begin{eqnarray}
L_n=\left\{ ( L^{\frac{n}{N-1}} )_{\ge -k+1};L \right\}_k=
\left\{ L( L^{\frac{n}{N-1}-1} )_{\ge-k+1} + R_n ; L \right\}_k
  \nonumber\\
 = LL_{n-(N-1)} +\{R_n;L\}_k , 
\end{eqnarray}

\noindent
and (\ref{SymmRel}) is satisfied. Evaluating powers of 
$(L( L^{\frac{n}{N-1}-1} )_{<-k+1})_{\ge -k +1} $ and
 $- (L( L^{\frac{n}{N-1}-1} )_{\ge -k+1})_{<-k+1} $  we get that 
$R_n$ has form (\ref{FormRem}).
$\Box$

\vspace{0.3cm}

\noindent
{\bf Lemma 3}. {\it A recursion  operator for the hierarchy  (\ref{Laxeqn})
is given by equalities, for $m=N-2, N-3, \dots , -1$, 

\begin{equation}
 \label{formulaS}
\matrix{
S_{m,n+(N-1)}=  \sum_{j=-1}^{m+1} S_jS_{m-j,n} \; +  
  \sum_{j=-1}^{m+1}(j+1-k) A_{j+1} S_{m-j,x} \; - \cr
  \sum_{j=-1}^{m+1} (m-j)A_{j+1,x} S_{m-j}, \cr
}
\end{equation}

\noindent
where to simplify the above formula  we  have defined that
 $S_{N-1}=1$ and $S_{N-1,x}=0$, $S_{N-1,n}=0$.
 Coefficients $ A_{N-2},A_{N-3},\dots, A_0 $ can be found
from the recursion relations, for $ m=N-2, \dots, -1$

\begin{eqnarray}
 \label{formulaA}
(N-1)A_{m,x}=\sum_{j=m}^{N-1} S_jS_{N-2+m-j} + 
 \sum_{j=m}^{N-2} (j+1-k)A_{j+1}S_{N-2+m-j,x} - \nonumber \\
   \sum_{j=m}^{N-2}(N-2+m-j)A_{j+1,x}S_{N-2+m-j} 
\end{eqnarray}
}

\vspace{0.3cm}

\noindent
{\bf Proof.}
Let us write the equality (\ref{SymmRel}), using (\ref{FormRem}) for $R_n$

$$
\sum_{i=-1}^{N-2} p^iS_{i,n+(N-1)} =
\left ( p^{N-1} + \sum_{i=-1}^{N-2} p^iS_{i}\right )
 \left ( \sum_{i=-1}^{N-2} p^iS_{i,n} \right ) +  
$$
$$   
p^k\left ( \sum_{j=0}^{N-1} (j-k)p^{j-k-1} A_j \right )
 \left ( \sum_{j=-1}^{N-2} p^{j} S_{j,x}  \right )  
$$
$$
-p^k\left ( \sum_{j=0}^{N-1} p^{j-k} A_{j,x}  \right )
  \left ( (N-1)p^{N-2}+\sum_{j=-1}^{N-2} jp^{j-1} S_j \right )
$$

\noindent
To have the equality, the coefficients of $p^{2N-3},\dots, p^{N-1}$ and $p^{-2}$
 must be zero, it gives  recursion relations to find 
$A_{N-2},\dots, A_0$.
The coefficients of $ p^{N-2},\dots, p^{-1}$
give the expressions for $S_{N-2,n+(N-1)},\dots, S_{-1,n+(N-1)}$.
$\Box$

\medskip

Although the recursion operator ${\cal R}$, given by (\ref{formulaS}), is a 
pseudo-differential operator, 
but it gives a hierarchy of local symmetries starting from the equation itself.
Indeed, equalities (\ref{formulaS}), (\ref{formulaA}) give expressions 
$S_{N-2,n+(N-1)}$ , $\dots$, $S_{-1,n+(N-1)}$ in terms
 of $S_{N-2}$ ,$\dots$, $ S_{-1}$  and $S_{N-2,n}$, $\dots$ , $S_{-1,n}$.
Hence, the recursion operator ${\cal R}$ is constructed in such a way that

\begin{equation}
\left\{ ( L^{\frac{n}{N-1}+1} )_{\ge -k+1};L \right\}_k=
{\cal R}\left(\left\{ ( L^{\frac{n}{N-1}} )_{\ge -k+1};L \right\}_k\right)
\end{equation}

\section*{IV. Some Integrable Systems}

We shall consider first some examples for $k=0$, $k=1$ and the 
general case in the next section.

\subsection*{A.  Multicomponent hierarchy containing also the shallow water 
wave equations , $k=0$}

This  hierarchy corresponds to the case $k=0$. Let us give the first equation
 of hierarchy and a recursion operator for $N=2,3$.

\vspace{0.3cm}

\noindent
{\bf Proposition 1}. {\it In the case $N=2$ one has the Lax function

$$L=p+S+P\,p^{-1}$$

\noindent
and the Lax equation for  $n=2$, given by (\ref{case0}), when $k=0$

\begin{equation}
\label{ShW2}
\matrix{
\frac{1}{2}S_t & = & SS_x+P_x , \cr
\frac{1}{2}P_t & = & SP_x+PS_x .\cr
}
\end{equation} 

\noindent
and the  recursion operator, given by (\ref{RecCase0}),

\begin{equation}
\label{ShW2Rec}
{\cal R}=\left(\matrix{
S + S_xD_x^{-1}& 2   \cr
 2P + P_xD^{-1}_x& S \cr
}
\right ).
\end{equation}
}

\vspace{0.3cm}

\noindent
These equations are known as the shallow water wave equations or
as the equations of polytropic gas dynamics for $\gamma=2$ (See 
Sec.VI). 

\noindent
The first two symmetries of the system (\ref{ShW2}) are given by

\begin{eqnarray}
S_{t_{1}}&=&(S^3 + 6SP)_x,  \nonumber \\
P_{t_{1}}&=&(3S^2P + 3P^2)_x,
\end{eqnarray}

\begin{eqnarray}
S_{t_{2}}&=&(S^4 + 12S^2P + 6P^2)_x, \nonumber \\
P_{t_{2}}&=&(4S^3P + 12SP^2)_x.
\end{eqnarray}

\noindent
These are all commuting symmetries.

\vspace{0.3cm}

\noindent
{\bf Remark 1.} {\it In symmetric variables the system (\ref{ShW2})  is
written as

\begin{equation}
\label{ShW2M}
\matrix{
\frac{1}{2}u_t & = & (u+v)u_x + uv_x,\cr
\frac{1}{2}v_t & = & vu_x + (u+v)v_x, \cr
} 
\end{equation}

\noindent
and the recursion operator (\ref{ShW2Rec}) takes the form

\begin{equation}
{\cal R}= \left( \matrix{
u+v + u_xD_x^{-1} & 2u + u_xD_x^{-1}  \cr
2v + v_xD_x^{-1} & u +v + v_xD_x^{-1}  \cr
}
\right ).
\end{equation}
}

\vspace{0.3cm}

\noindent
{\bf Proposition 2}.{\it  In the case  $N=3$ one has the Lax function

$$
L=p^2 + pS + P + p^{-1}Q
$$

\noindent
and the Lax equation with $n=3$ is

\begin{equation}
\label{ShW3}
\matrix{
\frac{1}{3}S_t & = &(\frac{1}{2} P
                     -\frac{1}{8}S^2)S_x + \frac{1}{2}SP_x +Q_x,\cr
\frac{1}{3}P_t & = & \frac{1}{2}QS_x
                     + (\frac{1}{8}S^2+\frac{1}{2}P)P_x + SQ_x,\cr
\frac{1}{3}Q_t & = & \frac{1}{4} SQS_x + \frac{1}{2}QP_x
                     + (\frac{1}{8}S^2 + \frac{1}{2}P)Q_x .\cr
}
\end{equation}

\noindent
The recursion operator, corresponding to this equation, is
\begin{equation}
 \label{ShW3Rec}
{\cal R}=\left ( \matrix{
-\frac{S^2}{4}+P+P_xD_x^{-1} -\frac{S_x}{4}D_x^{-1}\cdot S &
\frac{S}{2} +\frac{S_x}{2}D_x^{-1} & 3 \cr
\frac{3Q}{2}+(Q_x+ \frac{P_xS}{2})D_x^{-1} - \frac{P_x}{4}D_x^{-1}\cdot S &
          P +\frac{P_x}{2}D_x^{-1} & 2S \cr
\frac{SQ}{4}+(\frac{SQ_x}{2} + \frac{S_xQ}{2})D_x^{-1} -
                                           \frac{Q_x}{4}D_x^{-1}\cdot S &
\frac{3Q}{2} +\frac{Q_x}{2}D_x^{-1} & P \cr
}
\right ).
\end{equation}
}

\vspace{0.3cm}

\noindent
{\bf Proof.}
Using  (\ref{formulaA}) we find the function $R_n$ and using (\ref{formulaS}) we find the
recursion operator (\ref{ShW3Rec}). $\Box$

\vspace{0.3cm}

\noindent
{\bf Remark 2.} {\it In symmetric variables the equation (\ref{ShW3})  
is written as

\begin{equation}
\label{ShW3M}
\matrix{
\frac{1}{3}u_t& = &( -\frac{1}{8}u^2 + \frac{1}{2}(uv+uw+vw) 
                                            + \frac{1}{8}(v+w)^2 )u_x \cr
              &    & +(\frac{1}{4}u^2 + \frac{1}{4}uv + \frac{3}{4}uw)v_x +
                   (\frac{1}{4}u^2 + \frac{1}{4}uw + \frac{3}{4}uv)w_x ,\cr
\frac{1}{3}v_t& = &  (\frac{1}{4}v^2 + \frac{1}{4}uv + \frac{3}{4}vw)u_x
                     +(\frac{1}{4}v^2 + \frac{1}{4}vw + \frac{3}{4}uv)w_x \cr
              &   & + ( -\frac{1}{8}v^2 + \frac{1}{2}(uv+uw+vw) +
                                               \frac{1}{8}(u+w)^2 )v_x, \cr
\frac{1}{3}w_t& = & (\frac{1}{4}w^2 + \frac{1}{4}uw + \frac{3}{4}wv)u_x +
                   (\frac{1}{4}w^2 + \frac{1}{4}wv + \frac{3}{4}uw)v_x \cr
              &   & +( -\frac{1}{8}w^2 + \frac{1}{2}(uv+uw+vw)  + 
                                               \frac{1}{8}(v+u)^2 )w_x ,\cr
}
\end{equation}

\noindent
and the recursion operator takes the form (\ref{ShW3MRec}) 
given in the Appendix.}

\subsection*{B. Toda hierarchy \,($k=1$)}

Toda hierarchy corresponds to the case $k=1$\, \cite{Str}. 
Let us give the first equation of
hierarchy and a recursion  operator for $N=2$ and $N=3$.

\vspace{0.3cm}

\noindent
{\bf Proposition 3}. {\it In the case $N=2$ and $n=1$ one has the Lax function

$$L=p+S+P\,p^{-1}$$

\noindent
and the Lax equation for  $n=1$ , given by (\ref{case1}),

\begin{equation}
\label{Toda2}
\matrix{
S_t & = & P_x ,   \cr
P_t & = & PS_x , \cr
}
\end{equation}
 
\noindent
and the recursion operator, given by (\ref{RecCase1}),

\begin{equation}
\label{Toda2Rec}
{\cal R}=\left( \matrix{
S & 2 + P_xD_x^{-1}\cdot P^{-1}  \cr
 2P &  S + S_xPD_x^{-1}\cdot P^{-1} \cr
}
\right ).
\end{equation}
}

\vspace{0.3cm}

\noindent
The first two symmetries of the equation (\ref{Toda2}) are given by

\begin{eqnarray}
S_{t_{1}}&=&(2SP)_x,  \nonumber \\
P_{t_{1}}&=&P(2P+S^2)_x, 
\end{eqnarray}

\begin{eqnarray}
S_{t_{2}}&=&(3S^2P + 3P^2)_x, \nonumber \\
P_{t_{2}}&=&P(6PS + S^3)_x. 
\end{eqnarray}

\vspace{0.3cm}

\noindent
{\bf Remark 3.} {\it In symmetric variables the equation (\ref{Toda2})  is written
as

\begin{equation}
\matrix{
u_t & = & uv_x,  \cr
v_t & = & vu_x, \cr
}
\end{equation} 

\noindent
and the recursion operator (\ref{Toda2Rec}) takes the form

\begin{equation}
{\cal R}= \left( \matrix{
u+v + uv_xD_x^{-1}\cdot u^{-1} & 2u + uv_xD_x^{-1}\cdot v^{-1}   \cr
2v + vu_xD_x^{-1}\cdot u^{-1} & u +v + vu_xD_x^{-1}\cdot v^{-1}   \cr
}
\right ).
\end{equation}
}
\vspace{0.3cm}

\noindent
{\bf Proposition 4}. {\it In the case  $N=3$ and $n=1$ one has the Lax 
function 

$$
L=p^2 + pS_1 + P + p^{-1}Q
$$

\noindent
and the Lax equation with $n=1$ is

\begin{equation}
 \label{Toda3}
\matrix{
S_t & = & P_x -\frac{1}{2}SS_x,  \cr
P_t & = & Q_x,                    \cr
Q_t & = & \frac{1}{2} QS_x .     \cr
}
\end{equation} 

\noindent
The recursion operator, corresponding to this equation, is

\begin{equation}
 \label{Toda3Rec}
{\cal R}=\left ( \matrix {
 P - \frac{1}{4}S^2 + (\frac{1}{2} P_x - \frac{1}{4}SS_x)D_x^{-1} &
\frac{1}{2}S & 3 + 2Q_xD_x^{-1} \cdot Q^{-1} \cr
\frac{3}{2}Q +\frac{1}{2}Q_xD_x^{-1} & P & 2S + (SQ)_xD_x^{-1}\cdot Q^{-1}\cr
\frac{1}{4}SQ +\frac{1}{4}S_xQD_x^{-1} & \frac{3}{2}Q &
 P + P_xQD_x^{-1}\cdot Q^{-1}\cr
}
\right ).
\end{equation}

}

\vspace{0.3cm}

\noindent
{\bf Proof.}
Using equalities (\ref{formulaA}) we find  the function $R_n$ and
using (\ref{formulaS}) we find the recursion operator~(\ref{Toda3Rec}). $\Box$

\vspace{0.3cm}

\noindent
{ \bf Remark 4.} {\it In symmetric variables the equation (\ref{Toda3})  is 
written as

\begin{equation}
 \label{Toda3M}
\matrix{
u_t & = & \frac{1}{2}u (-u_x + v_x + w_x),\cr
v_t & = & \frac{1}{2}v (+u_x - v_x + w_x),\cr
w_t & = & \frac{1}{2}w (+u_x + v_x - w_x),\cr
}
\end{equation}
 
\noindent
and the recursion operator takes the form (\ref{Toda3MRec}) given in
the Appendix.} 

\section*{V. Lax equation for general $k$}

We shall only consider the case where $N=2$. We have the Lax function

\begin{equation}
L= p + S + Pp^{-1} 
\end{equation}

\noindent
and the Lax equation

\begin{equation}
\frac{\partial L}{\partial t_n}=
  \left\{ ( L^n )_{\ge -k+1};L \right\}_k.
\end{equation}

\noindent
We consider two cases $k\ge 1$ and $k\le 0$.

\subsection*{A. The first case $k \ge 1$ }

\noindent
{\bf Proposition 5}. {\it In the case $N=2$ and $k\ge 1$ one has the Lax 
equation

\begin{equation}
 \label{case1}
\matrix{
S_t&=&kP^{k-1}P_x, \cr
P_t&=&kP^{k}S_x. \cr
}
\end{equation}

\noindent
and the recursion operator for this equation is

\begin{equation}
\label{RecCase1}
{\cal R}=\left( \matrix {
S  + (1-k)S_x D_x^{-1} & 2 +kP^{k-1}P_xD_x^{-1}\cdot P^{-k} \cr
 2P +(1-k)P_xD_x^{-1} &  S + kS_xP^kD_x^{-1}\cdot P^{-k} }
\right ).
\end{equation}
}

\noindent
{\bf Proof.}
The smallest power of $p$ in $L^n$ is $-n$. To have
powers less than $-k+1$ we must put $n=k$. If there are no such powers
 then Poisson brackets are $ \left\{ ( L^n );L \right\}_k=0$.

Let us calculate the Lax equation

$$
L_t=\left\{ ( L^k )_{\ge -k+1};L \right\}_k=
-\left\{ ( L^k )_{\le -k};L \right\}_k
$$

\noindent
We have $( L^k )_{\le -k}=[(p+S+Pp^{-1})^k]_{\le -k}=P^kp^{-k}$, thus

$$
L_t=-\left\{ P^kp^{-k};p+S+Pp^{-1}\right\}_k .
$$

\noindent
And  we get the equation (\ref{case1}).
Using (\ref{formulaS}), (\ref{formulaA}) we  find the recursion 
operator (\ref{RecCase1}).
$\Box$

\vspace{0.3cm}

\noindent
First two symmetries are given as follows

\begin{eqnarray}
S_{t_{1}}&=&(P^{k}\,S)_{x}, \nonumber \\
P_{t_{1}}&=&P^{k}\,\left (P+{k \over 2}S^2 \right )_{x}, 
\end{eqnarray}

\begin{eqnarray}
S_{t_{2}}&=&(k+1)(k+2) \left ({1 \over 2}P^{k}S^2+{1 \over k+1}P^{k+1}
\right )_{x}, 
\nonumber \\
P_{t_{2}}&=&(k+1)(k+2)P^{k} \left (PS+{k \over 6}S^3 \right )_{x}. 
\end{eqnarray}

\vspace{0.3cm}

\noindent
{ \bf Remark 5.} {\it In symmetric variables the equation  (\ref{case1}) is written
as

\begin{equation}
\matrix{
u_t & = & ku^kv^{k-1}v_x, \cr
v_t & = & ku^{k-1}v^ku_x,  \cr
}
\end{equation} 

\noindent
and the recursion operator  (\ref{RecCase1}) takes the form

\begin{equation}
{\cal R}= \left(
\matrix{
u+v +(1-k)u_xD_x^{-1} + &  2u + (1-k)u_xD_x^{-1} + \cr
ku^kv^{k-1}v_xD_x^{-1}\cdot u^{-k}v^{-k+1} &
                       ku^kv^{k-1}v_xD_x^{-1}\cdot u^{-k+1}v^{-k}   \cr
 & \cr
2v + (1-k)v_xD_x^{-1} + & u +v +(1-k)v_xD_x^{-1} + \cr
 ku^{k-1}v^ku_xD_x^{-1}\cdot u^{-k}v^{-k+1} &
    ku^{k-1}v^ku_xD_x^{-1}\cdot u^{-k+1}v^{-k}   \cr
}
\right ).
\end{equation}
}

\subsection*{B. The  second case $k \le 0$ }

\noindent
{\bf Proposition 6}.  {\it In the case $N=2$ and $k\le 0$ one has the Lax 
equation 

\begin{equation}
 \label{case0}
\matrix{
S_t&=&(-k+2)(-k+1)SS_x + (-k+2)P_x, \cr
P_t&=&(-k+2)(-k+1)SP_x  + (-k+2)S_xP.\cr
}
\end{equation}

\noindent
and the recursion operator for this equation is

\begin{equation}
 \label{RecCase0}
{\cal R}=\left( \matrix{
S  + (1-k)S_x D_x^{-1} & 2 +kP^{k-1}P_xD_x^{-1}\cdot P^{-k}  \cr
 2P +(1-k)P_xD_x^{-1}&  S + kS_xP^kD_x^{-1}\cdot P^{-k} \cr
}
\right ).
\end{equation}
}

\vspace{0.3cm}

\noindent
{\bf Proof.}
The largest  power of $p$ in $L^n$ is $p^{n}$. To have
powers larger than $-k+1$ we must put $n=-k+1$. Then we have 

$$
( L^{-k+1} )_{\ge -k+1}=[(p+S+Pp^{-1})^{-k+1}]_{\ge -k+1}
= p^{-k+1},
$$

\noindent
 thus

$$
L_t=\left\{ p^{-k+1};p+S+Pp^{-1}\right\}_k .
$$

\noindent
Then the Lax equation becomes 

\begin{eqnarray*}
S_t&=&S_x,\\
P_t&=&P_x .
\end{eqnarray*}

\noindent
This is  a trivial equation,
let us calculate the second symmetry.
We have $( L^{-k+2} )_{\ge -k+1}=[(p+S+Pp^{-1})^{-k+1}]_{\ge -k+1}
= p^{-k+2} + (-k+2)Sp^{-k+1} $, thus

$$
L_t=\left\{ p^{-k+2} +  (-k+2)Sp^{-k+1};p+S+Pp^{-1}\right\}_k .
$$

\noindent
we get the equation (\ref{case0}).
Using (\ref{formulaS}), (\ref{formulaA}) we find the recursion 
operator (\ref{RecCase0}). $\Box$

\vspace{0.3cm}

\noindent
First two symmetries are given as follows

\begin{eqnarray}
S_{t_{1}}&=&(k-2)(k-3) \left (P\,S +\frac{1}{6}(1-k)S^3 \right 
)_{x},\nonumber\\
P_{t_{1}}&=&(k-2)(k-3) \left (SS_xP + \frac{1}{2}(1-k)S^2\, P_x + PP_x 
\right ), 
\end{eqnarray}

\begin{eqnarray}
S_{t_{2}}&=&(2-k)(3-k)(4-k) \left (\frac{1}{2}S^2P +\frac{1}{6}S^4 +
\frac{1}{2(2-k)}P^2 \right )_x ,\nonumber \\
P_{t_{2}}&=&(2-k)(3-k)(4-k) \left ( \frac{1}{2}S^2S_xP + 
\frac{1}{6}(1-k)S^3\, P_x  \right.
 \nonumber \\
&& \left. + SPP_x  + \frac{1}{(2-k)}P^2S_x \right ) . 
\end{eqnarray}

\vspace{0.3cm}
 
\noindent
{\bf Remark 6.} {\it In symmetric variables the equation (\ref{case0}) is written as

\begin{equation}
\matrix{
 u_t & = & (-k+2)(1-k)(u+v)u_x + (-k+2)uv_x, \cr
 v_t & = & (-k+2)vu_x + (-k+2)(1-k)(u+v)v_x,  \cr
}
\end{equation}

\noindent
and the recursion operator (\ref{RecCase0}) takes the form

\begin{equation}
{\cal R}=\left( \matrix{
u+v +(1-k)u_xD_x^{-1} + &  2u + (1-k)u_xD_x^{-1} + \cr
ku^kv^{k-1}v_xD_x^{-1}\cdot u^{-k}v^{-k+1} &
                       ku^kv^{k-1}v_xD_x^{-1}\cdot u^{-k+1}v^{-k}   \cr
 & \cr
2v + (1-k)v_xD_x^{-1} + & u +v +(1-k)v_xD_x^{-1} + \cr
 ku^{k-1}v^ku_xD_x^{-1}\cdot u^{-k}v^{-k+1} &
    ku^{k-1}v^ku_xD_x^{-1}\cdot u^{-k+1}v^{-k}   \cr
}
\right ).
\end{equation}
}

\vspace{0.3cm}

In this Section, to obtain the recursion operators we have considered
two different cases $k \le 0$ and $k \ge 1$ to simplify some technical 
problems in the method. At the end we obtained recursion operators having 
the same forms (\ref{RecCase1}) and (\ref{RecCase0}). Hence any one of these
represent the recursion operator for $k \in {\Bbb Z}$. It seems , comparing 
the results, that the systems
of equations in one case  are symmetries of the other case. For instance,
the system (\ref{case0}) is a symmetry of system (\ref{case1}). Hence 
we may consider only one case with recursion operator $(\ref{RecCase1})$
for all integer values of $k$.

\section*{VI. Lax function for polytropic gas dynamics}

In this section we consider another Lax function,  introduced in \cite{BurD},

\begin{equation}
\label{Laxfnc2}
L=p^{\gamma-1}+u+
 \frac{v^{\gamma-1}}{(\gamma-1)^2}p^{-\gamma+1}
\end{equation}

\noindent
and the Lax equation

\begin{equation}
\label{Laxeqn2}
\frac{\partial L}{\partial t}=\frac{\gamma-1}{\gamma}
\left \{ (L^{\frac{\gamma}{\gamma-1}})_{\ge 1},L \right \}_0 ,
\end{equation}

\noindent
gives the equations of the polytropic gas dynamics.

\vspace{0.3cm}

\noindent
{\bf Proposition 7}. {\it The Lax equation corresponding to (\ref{Laxeqn2}) is

\begin{equation}
\label{Gaseqn}
\matrix{
u_t+uu_x + v^{\gamma - 2}v_x&=&0, \cr
v_t + (uv)_x&=&0.\cr
}
\end{equation}
}
\vspace{0.3cm}

\noindent
{\bf Proof.}
 Expanding the function (\ref{Laxfnc2}) around the point $p=\infty$, we have

$$
\left ( p^{\gamma-1} + u + \frac{v^{\gamma-1}}{(\gamma-1)^2}p^{-\gamma+1} 
\right )^{\frac{\gamma}{\gamma-1}}=\\
p^\gamma +\frac{\gamma}{\gamma-1}pu + \dots
$$

\noindent
all other terms have negative powers of $p$.
Therefore

$$
\left( L^{\frac{\gamma}{\gamma-1}}\right )_{\ge 1}
=p^\gamma +\frac{\gamma}{\gamma-1}pu
$$

\noindent
and the Lax equation (\ref{Laxeqn2}) corresponds to  (\ref{Gaseqn}).
$\Box$

\vspace{0.3cm}

\noindent
{\bf Proposition 8}. {\it The recursion operator for the equation 
(\ref{Gaseqn}) is

\begin{equation}
\label{GasRec}
{\cal R}=\left ( \matrix{
u+\frac {u_x}{\gamma-1}D_x^{-1} &
\frac{2v^{\gamma-2}}{\gamma-1} + \frac{(v^{\gamma-2})_x}{\gamma-1}D_x^{-1}\cr
 \frac{2v}{\gamma-1} + \frac{v_x}{\gamma-1}D_x^{-1}&
 u + \frac{\gamma-2}{\gamma-1}u_xD_x^{-1}\cr
}
\right ).
\end{equation}
}

\vspace{0.3cm}

\noindent
{\bf Proof.} Using the equation

$$
\frac{\partial L}{\partial t_{n+1}}=
L\frac{\partial L}{\partial t_n} +\{R_n,L\}. 
$$

\noindent
in the same way as for polynomial Lax function one can find the
recursion operator (\ref{GasRec}).
$\Box$

\noindent
It is interesting to note that the equation (\ref{case0})
and equations of polytropic gas dynamics (\ref{Gaseqn})
are related by the following change of variables

\begin{equation}
\label{GaseqnRec}
\matrix{
S= & \displaystyle \frac{u}{(-k+2)(-k+1)},\cr
&  \cr
P= & \displaystyle \frac{v^{\frac{1}{-k+1}}}{(-k+2)^2},\cr
}
\end{equation}

\noindent
where $\gamma=\displaystyle \frac{-k+2}{-k+1}$.
We note that under this change  of variables recursion operator 
(\ref{RecCase0}) is mapped to the  recursion operator (\ref{GasRec}).

\section*{VII. Reduction}

We consider reduction of the equation (\ref{Laxeqn}), written in
symmetric variables, by
setting $u_{1}=0$. Let us write the equation (\ref{Laxeqn}) as

\begin{equation}
\label{RedGen}
\Delta (u_N, \dots ,u_{1})=0,
\end{equation}

\noindent
where $\Delta$ is a differential operator. Then

\begin{equation}
\label{Reduced}
\Delta(u_N,\dots ,u_1) |_{u_{1}=0} = \left (
\begin{tabular}{c}
$\tilde \Delta ( u_N, \dots ,u_2),$ \\
\hline
  $ 0  $     \\
\end{tabular} \right )
\end{equation}

\noindent
where $\tilde \Delta$ is another differential operator.
Indeed, following \cite{StrF} for the  Lax function 
$
L= \frac{1}{p}\prod_{j=1}^{N}(p-u_j)
$
we have

$$
\frac{\partial L}{\partial t}=L\sum_{j=1}^{N}\frac{u_{j,t}}{p+u_j},
$$

$$
\frac{\partial L}{\partial x}=L\sum_{j=1}^{N}\frac{u_{j,x}}{p+u_j}
$$

\noindent
and

$$
\frac{\partial L}{\partial p}=L\left(- \frac{1}{p} +
  \sum_{j=1}^{N}\frac{1}{p+u_j} \right).
$$

\noindent
Thus  $u_{j,t}=Res_{p=-u_j} \{M,L\}_k$, where 
 $M=(L^{\frac{n}{N-1}})_{\ge -k+1}$.
The Lax equation (\ref{Laxeqn}) can be written as

$$
\sum_{j=1}^{N}\frac{u_{j,t}}{p+u_j}=p^kM_p\sum_{j=1}^{N}\frac{u_{j,x}}{p+u_j}
- p^kM_x \left( -\frac{1}{p} + \sum_{j=1}^{N}\frac{1}{p+u_j} \right).
$$

\noindent
Note, that $p^kM_x$ and $p^kM_p$ are polynomials. 
So, if we put $u_1=0$ and calculate residue  of right hand side at $p=0$ 
we get (\ref{Reduced}).  
A new equation 

\begin{equation}
 \label{NewEqn}
\tilde\Delta(u_N, \dots ,u_2)=0 
\end{equation}

\noindent
 is also integrable
and a recursion operator of this equation can be obtained as reduction of the 
recursion operator of the equation (\ref{RedGen}).
Let ${\cal R}$ be the recursion operator of (\ref{RedGen}) given  
by lemma 3 , then

\begin{equation}
\label{ReducedRec}
{\cal R} |_{u_{1}=0}= \left(
\begin{tabular}{c|c}
$\tilde R $ & $*$ \\
\hline
$0 \dots 0$ & $ 0 $\\
\end{tabular} \right).
\end{equation}  

\noindent
Indeed, we found the recursion operator using formula (\ref{SymmRel}).
This formula can be written as

$$
\sum_{j=1}^{N}\frac{u_{j,t_n}}{p+u_j}=
LL_{n-(N-1)} + 
p^kR_{n,p}\sum_{j=1}^{N}\frac{u_{j,x}}{p+u_j}
- p^kR_{n,x} \left( -\frac{1}{p} + \sum_{j=1}^{N}\frac{1}{p+u_j} \right)
$$

\noindent
and in the same way as for reduction  of (\ref{RedGen}) we have  (\ref{ReducedRec}), note, that 
$p^kR_{n,x}$  and $p^kR_{n,p}$ are also polynomials. 

\vspace{0.3cm}

\noindent
{\bf Lemma 4}. {\it The operator $\tilde R$ is a recursion operator of 
the equation (\ref{NewEqn}).}

\vspace{0.3cm}

\noindent
{\bf Proof.} Equation (\ref{NewEqn}) is an evolution equation, so, to 
prove that
$\tilde R$ is a recursion operator we must prove that for any solution 
$(u_N,\dots, u_2)$ of (\ref{NewEqn}) the following equality holds
(see \cite{Olv} )

$$
D_{\tilde\Delta} \tilde R =\tilde R  D_{\tilde\Delta},
$$

\noindent
where $ D_{\tilde\Delta} $ is a Frechet derivative of $ \tilde\Delta $.

If $(u_N,\dots, u_2)$ is a solution of (\ref{NewEqn}) then 
$(u_N,\dots, u_2,u_1=0)$ is a solution  of (\ref{RedGen}) and for the solution
$(u_N,\dots, u_2,u_1=0)$ we have

\begin{equation}
\label{SymmEql}
D_{\Delta}\, {\cal R} = {\cal R}\, D_{\Delta}. 
\end{equation}

\noindent
Next

$$
D_{\Delta}|_{u_1=0}=\left(
\begin{tabular}{c|c}
$\tilde D $ & $ *$ \\
\hline
$ 0 \dots 0 $&  $ * $ \\
\end{tabular} \right)
$$

\noindent
and

$$
{\cal R}|_{u_1=0}=\left(
\begin{tabular}{c|c}
$\tilde R $  & $ * $\\
\hline
$ 0 \dots 0 $ & $ 0  $ \\
\end{tabular} \right).
$$

\noindent
Hence by (\ref{SymmEql}) we have $ \tilde D \tilde R =\tilde R \tilde D $.
Calculating Frechet derivative we take derivatives with respect to one
variable,
considering other variables as constants. Thus,  to calculate $\tilde D$
we can put $u_1=0$ and differentiate with respect to other variables or
we can first differentiate and then put  $u_1=0$.
It means that $\tilde D=D_{\tilde\Delta}$ and 

$$
 D_{\tilde\Delta} \tilde R =\tilde R  D_{\tilde\Delta}.
$$
$\Box$

\medskip

Let us consider reduction of systems, given by remark 2  and
remark 4 and their recursion operators.

\vspace{0.3cm}

\noindent
{\bf Proposition 9}. {\it
Putting $w=0$ in  (\ref{Toda3M}) and (\ref{Toda3MRec}) we obtain a new system

\begin{equation}
\matrix{
u_t&=&\frac{1}{2}u (-u_x + v_x ),\cr
v_t&=&\frac{1}{2}v (+u_x - v_x),\cr
 } 
\end{equation}

\noindent
and its recursion operator

\begin{equation}
{\cal R}=\left ( \matrix{
-uv+\frac{u}{4}(u+v)         & -\frac{u}{4}(u+v) \cr
+\frac{u}{4}(u_x-v_x)D_x^{-1}& + \frac{u}{4}(u_x-v_x)D_x^{-1}\cr
 & \cr
-\frac{v}{4}(u+v) & -uv +\frac{v}{4}(u+v) \cr
+ \frac{v}{4}(-u_x+v_x)D_x^{-1}& +\frac{v}{4}(-u_x+v_x)D_x^{-1} \cr
}
\right )
\end{equation}

\noindent
respectively.
}
$\Box$

\vspace{0.3cm}

\noindent
{\bf Proposition 10}. {\it
 Putting $w=0$ in  (\ref{ShW3M}) and (\ref{ShW3MRec}) we obtain a new system 

\begin{equation}
\label{ShW3MRed}
\matrix{
\frac{1}{3}u_t& = &( -\frac{1}{8}u^2 + \frac{1}{2}uv  + \frac{1}{8}v^2 )u_x
                                             +(\frac{1}{4}u^2 + \frac{1}{4}uv )v_x, \cr
\frac{1}{3}v_t& = &  (\frac{1}{4}v^2 + \frac{1}{4}uv )u_x 
                   + ( -\frac{1}{8}v^2 + \frac{1}{2}uv + \frac{1}{8}u^2 )v_x,\cr 
}
\end{equation}

\noindent
and its recursion operator

\begin{equation}
{\cal R}=\left( \matrix {
-\frac{u^2}{4} + \frac{3uv}{4} + (\frac{u_xv}{2} +\frac{uv_x}{2})D_x^{-1} & 
\frac{u}{4}(u+v) + (\frac{u_xv}{2} +\frac{uv_x}{2})D_x^{-1}  \cr
-\frac{u_x}{4}D_x^{-1}\cdot u + \frac{u_x}{4}D_x^{-1}\cdot v &
+\frac{u_x}{4}D_x^{-1}\cdot u - \frac{u_x}{4}D_x^{-1}\cdot v\cr
 &  \cr
\frac{v}{4}(u+v) + (\frac{uv_x}{2} +\frac{u_xv}{2})D_x^{-1} & 
-\frac{v^2}{4} + \frac{3uv}{4} +  (\frac{uv_x}{2} +\frac{u_xv}{2})D_x^{-1}    \cr                 
-\frac{v_x}{4}D_x^{-1}\cdot u + \frac{v_x}{4}D_x^{-1}\cdot v &
+\frac{v_x}{4}D_x^{-1}\cdot u - \frac{v_x}{4}D_x^{-1}\cdot v \cr
}
 \right ),
\end{equation}

\noindent
respectively.
}
$\Box$

It is worth to mention that  by reduction  we obtain a new equation.
For example, consider the case $k=0$.
The  equation (\ref{ShW2M}, corresponding to  $N=2$, and reduction of the
equation (\ref{ShW3M}), corresponding to  $N=3$, are not related by
 a linear transformation of variables. Indeed, in the equation (\ref{ShW2M}))
coefficients of $u_x, v_x$ are linear in $u,v$ but in the
 equation (\ref{ShW3MRed})  coefficients of $u_x, v_x$ contain
 quadratic terms. Hence they can not be related by a linear transformation.

\section*{VII. Conclusion}

We have constructed the recursion operators of some equations of 
hydrodynamic type. The form of the these operators fall into the
class of pseudo differential operators $A+B\,D^{-1}$ where
$A$ and $B$ are functions of dynamical variables and their derivatives.
The generalized symmetries of these equations are local and all
belong to the same class (i.e., they are also equations of 
hydrodynamic type). We have introduced a method of reduction which 
leads also to integrable class.
These properties, bi Hamiltonian structure of
the equations we obtained and equations with rational Lax functions
 will be communicated elsewhere.

\section*{Acknowledgments}

We thank Burak G{\" u}rel and Atalay Karasu for several discussions.
This work is partially supported by the Scientific and Technical
Research Council of Turkey and by Turkish Academy of Sciences.

\vspace{0.3cm}

\noindent
{\Large \bf Appendix.}  Recursion operators of the systems (\ref{ShW3M}) and
(\ref{Toda3M}) are respectively  given by

\begin{equation}
\label{ShW3MRec}
{\cal R}=\left( \matrix{
-\frac{u^2}{4} + \frac{3}{4}(uv+uw) + wv & \frac{u}{4}(u+v+w) + \frac{3uw}{2} &
                                                        \frac{u}{4}(u+v+w) + \frac{3uv}{2} \cr
+\frac{u_x}{2}(v+w)D_x^{-1}  & + \frac{u_x}{2}(v+w)D_x^{-1}  & +\frac{u_x}{2}(v+w)D_x^{-1} \cr
+\frac{u}{2}(v_x+w_x)D_x^{-1} & +\frac{u}{2}(v_x+w_x)D_x^{-1} & +\frac{u}{2}(v_x+w_x)D_x^{-1} \cr
-\frac{u_x}{4}D_x^{-1}\cdot u + \frac{u_x}{4}D_x^{-1}\cdot v &
+\frac{u_x}{4}D_x^{-1}\cdot u - \frac{u_x}{4}D_x^{-1}\cdot v&
+\frac{u_x}{4}D_x^{-1}\cdot u + \frac{u_x}{4}D_x^{-1}\cdot v \cr
 + \frac{u_x}{4}D_x^{-1}\cdot w &
+ \frac{u_x}{4}D_x^{-1}\cdot w &
- \frac{u_x}{4}D_x^{-1}\cdot w \cr
 &  &  \cr
\frac{v}{4}(u+v+w) + \frac{3vw}{2} & -\frac{v^2}{4} + \frac{3}{4}(uv+vw) + uw &
                                                               \frac{v}{4}(u+v+w) + \frac{3uv}{2} \cr
+\frac{v_x}{2}(u+w)D_x^{-1} & +\frac{v_x}{2}(u+w) D_x^{-1} & +\frac{v_x}{2}(u+w) D_x^{-1}\cr
+\frac{v}{2}(u_x+w_x)D_x^{-1} & +\frac{v}{2}(u_x+w_x)D_x^{-1} & +\frac{v}{2}(u_x+w_x)D_x^{-1}\cr
-\frac{v_x}{4}D_x^{-1}\cdot u + \frac{v_x}{4}D_x^{-1}\cdot v &
+\frac{v_x}{4}D_x^{-1}\cdot u - \frac{v_x}{4}D_x^{-1}\cdot v &
+\frac{v_x}{4}D_x^{-1}\cdot u + \frac{v_x}{4}D_x^{-1}\cdot v \cr
+ \frac{v_x}{4}D_x^{-1}\cdot w &
 + \frac{v_x}{4}D_x^{-1}\cdot w &
- \frac{v_x}{4}D_x^{-1}\cdot w \cr
&  &  \cr
\frac{w}{4}(u+v+w) + \frac{3vw}{2} & \frac{w}{4}(u+v+w) + \frac{3uw}{2} &
                                                              -\frac{w^2}{4} + \frac{3}{4}(uw+vw) + uv \cr
+\frac{w_x}{2}(u+v)D_x^{-1} &+\frac{w_x}{2}(u+v)D_x^{-1} & +\frac{w_x}{2}(u+v)D_x^{-1} \cr
+\frac{w}{2}(u_x+v_x)D_x^{-1} &+ \frac{w}{2}(u_x+v_x)D_x^{-1} & +\frac{w}{2}(u_x+v_x)D_x^{-1} \cr
-\frac{w_x}{4}D_x^{-1}\cdot u + \frac{w_x}{4}D_x^{-1}\cdot v &
+\frac{w_x}{4}D_x^{-1}\cdot u - \frac{w_x}{4}D_x^{-1}\cdot v &
+\frac{w_x}{4}D_x^{-1}\cdot u + \frac{w_x}{4}D_x^{-1}\cdot v \cr
+ \frac{w_x}{4}D_x^{-1}\cdot w &
+ \frac{w_x}{4}D_x^{-1}\cdot w &
- \frac{w_x}{4}D_x^{-1}\cdot w \cr
}\right ).
\end{equation}

\newpage

\begin{equation}
\label{Toda3MRec}
{\cal R}=\left ( \matrix{
-(uv+uw+vw) & -\frac{u}{4}(u+v+w) & -\frac{u}{4}(u+v+w) \cr
+\frac{u}{4}(u+v+w) & -\frac{3uw}{2} & -\frac{3uv}{2} \cr
+\frac{u}{4}(u_x-v_x-w_x)D_x^{-1}& + \frac{u}{4}(u_x-v_x-w_x)D_x^{-1}&
 + \frac{u}{4}(u_x-v_x-w_x)D_x^{-1}\cr
-u(wv_x +vw_x)D_x^{-1}\cdot u^{-1}& -u(wv_x +vw_x)D_x^{-1}\cdot v^{-1}&
-u(wv_x +vw_x)D_x^{-1}\cdot w^{-1}\cr
 & & \cr
-\frac{v}{4}(u+v+w) & -(uv+uw+vw) & -\frac{v}{4}(u+v+w) \cr
-\frac{3vw}{2} & +\frac{v}{4}(u+v+w) & -\frac{3uv}{2}\cr
+ \frac{v}{4}(-u_x+v_x-w_x)D_x^{-1}& +\frac{v}{4}(-u_x+v_x-w_x)D_x^{-1} &
+ \frac{v}{4}(-u_x+v_x-w_x)D_x^{-1}\cr
-v(wu_x +uw_x)D_x^{-1}\cdot u^{-1} & -v(wu_x +uw_x)D_x^{-1}\cdot v^{-1} &
-v(wu_x +uw_x)D_x^{-1}\cdot w^{-1}\cr
 & & \cr
-\frac{w}{4}(u+v+w) & -\frac{w}{4}(u+v+w) & -(uv+uw+vw)\cr
- \frac{3uw}{2} & -\frac{3vw}{2} & +\frac{w}{4}(u+v+w)\cr
+ \frac{w}{4}(-u_x-v_x+w_x)D_x^{-1} &  + \frac{w}{4}(-u_x-v_x+w_x)D_x^{-1} &
+\frac{w}{4}(-u_x-v_x+w_x)D_x^{-1}\cr
-w(uv_x +vu_x)D_x^{-1}\cdot u^{-1} & -w(uv_x +vu_x)D_x^{-1}\cdot v^{-1} &
-w(uv_x +vu_x)D_x^{-1}\cdot w^{-1}\cr
}\right ).
\end{equation}



\begin{thebibliography}{100}
\bibitem{GKS} M. G\"urses, A. Karasu,  V.V. Sokolov, " On construction
of recursion  operator from Lax representation", {\it J. Math. Phys},
{\bf 40}, 6473-6490 (1999).
\bibitem{GelDik} I.M. Gel'fand, L.A. Dikkii, Asymptotic  Behavior of the
 Re-solvent of Sturm-Liouville equations and the Algebra of the
Korteweg-de Vrise equations, {\it Funk. Anal. Appl.},{\bf 10},13 (1976).
\bibitem{gur2} Preliminary version of this work was first reported in
M. G{\" u}rses and K. Zhelthukin, "On Construction of Recursion Operators
for Some Equations of Hydrodynamic Type", in {\it International Conference
on Complex Analysis , Differential Equations and Related Topics},
May 29-June 3, 2000, Ufa, Russia.
\bibitem{nd} B.A. Dubrovin and S.P. Novikov, "Hamiltonian formalism
 of one-dimensional systems of hydrodynamic type",
 {\it Soviet Math. Dokl.} , {\bf27}, 665 (1983).
\bibitem{frpt} E.V. Ferepantov, "Hydrodynamic-type systems",
in {\it CRC Handbook of Lie Group Analysis
of Differential Equations}, Vol. 1. Ed. N.H. Ibragimov, p. 303-31. CRC Press, 
New York, 1994
\bibitem{Olv} P.J. Olver, {\it Applications of Lie Groups to Differential 
Equations}, Second Edition, Graduate Text in Mathematics, Vol. 107.
 Springer-Verlag, New-York (1993).
\bibitem{PG} A. Weinstein, "Poisson Geometry",
{\it Differential Geometry and its applications}, {\bf 9}, 213-238 (1998).
\bibitem{StrF} D.B. Fairlie and I.A.B. Strachan, "The algebraic and 
Hamiltonian structure of the dispersion-less  Benney and Toda
hierarchies", {\it Inverse Problems},{\bf 12}, 885-908 (1998). 
\bibitem{Str} I.A.B. Strachan, ~" Degenerate Frobenius manifolds and
the ~~bi-Hamiltonian structure of rational Lax equations", {\it
J. Math. Phys.}, {\bf 40}, 5058-5079 (1999). 
\bibitem{BurD} J.C. Brunelli, A. Das, " A Lax description for
polytropic gas dynamics", {\it Phys. Lett. A },235 , 597-602 (1997).
\bibitem{BGZ} J.C. Brunelli, M. G{\" u}rses, and K. Zhelthukin, "On the
integrability of some Monge-Ampere' equations", {\it
Reviews in Mathematical Physics}, 2000.
\bibitem{sheft} M.B. Sheftel, "Generalized Hydrodynamic-type systems"
in {\it CRC Handbook of Lie Group Analysis of Differential Equations} ,
Vol. 3. Edited by  N.H. Ibragimov, p.169-189. CRC Press , New York , 1996.
\bibitem{tes} V.M. Teshukov, ``Hyperbolic systems admitting a nontrivial
 Lie-Backlund group'', {\it LIIAN}, {\bf 106}, 25-30, 1989.
\bibitem{fg} A.P. Fordy and B. G{\" u}rel, "A new construction of
recursion operators for systems of hydrodynamic type",
{\it Theoret. Math. Phys.}, 1999.
\end{thebibliography}
\end{document}